\documentclass[journal,11pt,draftclsnofoot,onecolumn]{IEEEtran}

\usepackage{amsmath,graphicx}
\usepackage{latexsym}
\usepackage{amsfonts}
\usepackage{graphicx} 
\usepackage{subfigure}

\usepackage{amssymb}
\usepackage{natbib}
\usepackage{enumerate}
\usepackage{amsmath}
\usepackage[usenames]{color}
\usepackage{longtable} 
\usepackage{tabularx}
\usepackage{chngpage}
\usepackage{lipsum}
\usepackage{booktabs}
\usepackage{verbatim}

\usepackage{bm}

\def\BibTeX{{\rm B\kern-.05em{\sc i\kern-.025em b}\kern-.08em
    T\kern-.1667em\lower.7ex\hbox{E}\kern-.125emX}}

\title{Fully Adaptive Gaussian Mixture Metropolis-Hastings Algorithm}
\author{David Luengo$^\dagger$, Luca Martino$^\ddagger$\\
$^\dagger$Department of Circuits and Systems Engineering, Universidad Polit\'ecnica de Madrid.\\
Carretera de Valencia Km. 7, 28031 Madrid, Spain.\\
$^\ddagger$Department of Signal Theory and Communications, Universidad Carlos III de Madrid.\\
Avenida de la Universidad 30, 28911 Legan\'es, Madrid, Spain.\\
E-mail: {\tt luca.martino@uc3m.es, david.luengo@upm.es}}

\begin{document}
%
\maketitle
\begin{abstract}
Markov Chain Monte Carlo methods are widely used in signal processing and communications for statistical inference and stochastic optimization. In this work, we introduce an efficient adaptive Metropolis-Hastings algorithm to draw samples from generic multi-modal and multi-dimensional target distributions. The proposal density is a mixture of Gaussian densities with all parameters (weights, mean vectors and covariance matrices) updated using all the previously generated samples applying simple recursive rules. Numerical results for the one and two-dimensional cases are provided.
\end{abstract}
\section{Introduction}
\label{sec:intro}

Markov Chain Monte Carlo (MCMC) methods \citep{Liu04b,Liang10} are ubiquitously used for performing inference and solving optimization problems in many scientific fields: statistics, digital communications, machine learning, signal processing, etc. \citep{Robert04,Wang2002,Andrieu2003,Fitzgerald01}.
MCMC approaches are able to generate samples virtually from any target distribution (known up to a normalizing constant) by using a simpler proposal distribution.
The basic underlying idea of standard MCMC techniques is producing a Markov chain that converges to the target.

The most famous MCMC technique is the Metropolis-Hastings (MH) algorithm \citep{Metropolis53,Hastings70,Robert04}. 
However, the main drawback of the MH method (and in general of all MCMC methods) is that the correlation among the samples in the Markov chain can be very high when the acceptance rate is low \citep{Liu04b,Liang10,MartinoSigPro10}.
Correlated samples provide less statistical information and the resulting chain can remain trapped almost indefinitely in a local mode, meaning that convergence can be extremely slow.
Therefore, since the correlation depends on the discrepancy between the target and proposal distributions, we would like it to be as close to the target as possible.

Several extensions have been proposed in the literature to speed up the convergence and reduce the so called ``burn-in'' period.
Among them, adaptive MH methods (i.e., MH algorithms with adaptive proposal distributions) are particularly interesting \citep{Liang10,MartinoA2RMS}. 
Indeed, MCMC techniques usually need the selection of several parameters by the user before they can be applied to any particular problem.
The use of adaptive proposals overcomes this issue, providing \emph{black-box} algorithms with \emph{self-tuning} capabilities.
An adaptive MH technique improves the proposal distribution by learning at least some of its parameters from all the previously generated samples.
%
Unfortunately, an important problem with the adaptation of the proposal is that the Markov property is lost and the invariant distribution of the chain could be disturbed. Hence, adaptive MH algorithms must be carefully designed to avoid this issue.

An adaptive Metropolis that uses an adaptive random walk Gaussian proposal, was introduced in \citep{Haario01} (we will denote it as AM method).
The covariance matrix of the proposal is updated using recursive empirical estimators applied to the samples generated by the chain.
The AM algorithm is an example of a partially adaptive MH approach, since it only updates the covariance of the proposal, whereas the mean of the Gaussian jumps to the current state of the chain at each iteration.
An attempt of extending the AM algorithm by using a mixture of Gaussians as a proposal and updating all of its parameters (thus obtaining a fully adaptive MH algorithm) can be found in \citep{Giordani10}.
However, the resulting algorithm is quite complicated, and the proposal is only updated at some iterations.

In this work, we introduce an independent MH technique where the proposal PDF is an adaptive mixture of Gaussians.
All the parameters (weights, means and covariance matrices) of the Gaussians in the mixture are updated using empirical estimators with simple recursive formulas (i.e., our method is fully adaptive).
After a training period, the proposal is adapted at every iteration.
The resulting AGM-MH algorithm can be used to draw samples from arbitrary multi-modal and multi-dimensional targets, always improving the performance w.r.t. a non-adaptive MH scheme using the initial proposal.

The rest of the paper is organized as follows.
Section \ref{sec:problem} shows the problem formulation.
Section \ref{sec:algorithm} presents the proposed AGM-MH algorithm.
Sections \ref{sec:recursive} and \ref{sec:blackBox} describe efficient parameter update rules and black-box usage.
Finally, Sections \ref{sec:results} and \ref{sec:discussion} show the results and conclude the paper.

\section{Problem Formulation}
\label{sec:problem}
Let us assume that we need to draw samples from a (possibly multi-modal) generic $d$-dimensional target probability density function (PDF), $p_o({\bf x})$, with support $\mathcal{D} \subseteq \mathbb{R}^d$.
The AM algorithm \citep{Haario01} uses an adaptive random walk MH with a Gaussian proposal, mean equal to the previous state of the chain (${\bm \mu} = {\bf x}_{t-1}$), and covariance matrix, ${\bf C}$, estimated from all previous states, i.e., $q({\bf x}|{\bf x}_{t-1}, {\bf C}) \propto \mathcal{N}({\bf x}| {\bf x}_{t-1}, {\bf C})$, where
\begin{equation}
	\mathcal{N}({\bf x}| {\bm \mu}, {\bf C})
		\propto \exp\left(-\frac{1}{2}({\bf x}-{\bm \mu})^{\top} {\bf C}^{-1} ({\bf x}-{\bm \mu})\right)
\label{eq:gaussian}
\end{equation}
denotes a standard multi-variate Gaussian PDF.
%
%
In order to improve the performance of the AM algorithm, here we consider a mixture of $N$ Gaussians as the proposal PDF, i.e.,
\begin{equation}
	q({\bf x}|{\bf w}, {\bm \mu}_{1:N}, {\bf C}_{1:N}) = \sum_{i=1}^{N}{w_i \mathcal{N}({\bf x}| {\bm \mu}_i, {\bf C}_i)},
\label{eq:GaussianProposal}
\end{equation}
where $\mathcal{N}({\bf x}| {\bm \mu}_i, {\bf C}_i)$ is given by \eqref{eq:gaussian}, 
${\bm \mu}_i=[\mu_{i,1},...,\mu_{i,d}]^T$ is the $d \times 1$ mean vector, ${\bf C}_i$ is the $d \times d$ positive definite covariance matrix, and ${\bf w}=[w_1,...,w_N]^T$ are the normalized weights (i.e., $w_1+...+w_N=1$).
Moreover, we define ${\bm \mu}_{1:N}=[{\bm \mu}_1,...,{\bm \mu}_N]$ and  ${\bf C}_{1:N}=[{\bf C}_1,...,{\bf C}_N]$.
We note that our approach is a fully adaptive MH algorithm, since (unlike the AM algorithm, which only updates the covariance) all the parameters in the mixture are learnt from all the previously generated samples.
The resulting algorithm is very simple, since the adaptation is based on empirical estimators that can be implemented efficiently using recursive formulas.

Since the adaptation could disturb the convergence of the generated chain to the target PDF, we consider the possibility of stopping it at an iteration $T_{stop}$.
Hence, for $t>T_{stop}$ our algorithm is a standard MH with an improved proposal PDF w.r.t. the initial choice, thus providing a better performance and guaranteeing convergence.
However, the numerical results described in Section \ref{sec:results} show that the algorithm seems to maintain the correct ergodicity properties, so we always use $T_{stop}=T_{tot}$.
A theoretical convergence proof is under development and will be included in future works.  
Finally, we note that degeneracy problems can appear during the first iterations in the update of the covariance matrices if we have a poor initialization.
In order to avoid this issue, we allow the method to use a few iterations ($t=1,\ldots,T_{train}$) to collect information about the target, assigning the produced state of the chain to the closest Gaussian in the mixture, as in \citep{Haario01}.

\section{AGM-MH Algorithm}
\label{sec:algorithm}
\label{sec_Alg}

The proposed AGM-MH algorithm is described below.
%
First of all, notice that, during the first $T_{train}$ time steps the algorithm just assigns the current state ${\bf x}_t$ of the chain to a Gaussian among the $N$ in the mixture, according to the minimum Euclidean distance between ${\bf x}_t$ and the means ${\bm \mu}_i^{(t-1)}$, $i=1,...,N$.
Afterwards, the algorithm updates all the parameters in the mixture until $t=T_{stop}$, when adaptation is stopped.
In the description of the algorithm, the parameters are updated using a block procedure, but efficient recursive update formulas can be obtained, as shown in Section \ref{sec:recursive}. 

\begin{enumerate}
\item {\bf Initialization:} 
\begin{enumerate}
\item {\it Time instants:} Set $t=0$. Choose also the values ${\bf x}_0\in \mathcal{D}$,  $T_{train}<T_{tot}$ and $T_{train}<T_{stop}<T_{tot}$. Let be $T_{tot}$ the number of total iteration of the chain.
\item {\it Proposal:} Choose the number of Gaussians $N$, as well the initial settings for ${\bm \mu}_{1:N}^{(0)}=[{\bm \mu}_1^{(0)},...,{\bm \mu}_N^{(0)}]$ and ${\bf C}_{1:N}^{(0)}=[{\bf C}_1^{(0)},...,{\bf C}_N^{(0)}]$. Set ${\bf w}^{(0)}=\frac{1}{N}{\bf 1}$.
\item {\it Auxiliary parameters:} Define ${\bf S}_i^{(0)}\triangleq [{\bf s}_i^{(1)}={\bm \mu}_{i}^{(0)}]$, and $m_i=1$ represents the number of columns of ${\bf S}_i^{(0)}$, with $i=1,...,N$. Let $\epsilon$ a small constant value and ${\bf I}_d$ a identity matrix.
\end{enumerate}

\item {\bf MH steps:}
\begin{enumerate}
\item Sample ${\bf x}'$ from a mixture of Gaussian PDFs,
 $${\bf x}'\sim q_t({\bf x}' | {\bf w}^{(t)},{\bm \mu}_{1:N}^{(t)}, {\bf C}_{1:N}^{(t)} ).$$
\item Accept ${\bf x}_{t+1}={\bf x}'$ with probability
\begin{equation}
\small
\alpha=\min\left[1, \frac{p({\bf x}') q({\bf x}_{t}|{\bf w}^{(t)},{\bm \mu}_{1:N}^{(t)}, {\bf C}_{1:N}^{(t)})}{p({\bf x}_t) q({\bf x}'|{\bf w}^{(t)},{\bm \mu}_{1:N}^{(t)}, {\bf C}_{1:N}^{(t)})} \right],
\label{AlphaEq}
\end{equation}
otherwise set ${\bf x}_{t+1}={\bf x}_t$. 
\end{enumerate}

\item If $t<T_{stop}$, {\bf update parameters of the proposal:}
         \begin{enumerate}
                     \item Find the closest Gaussian to ${\bf x}_{t+1}$ (w.r.t. Euclidean distance), i.e., find the index
                            \begin{equation}
                                   j=\arg \min_{i} | {\bm \mu}_i^{(t)}- {\bf x}_{t+1}|^2.
                           \end{equation}
                     \item \label{Step3b}  Set $m_j=m_j+1$ and update (adding a new column) the $j$-th auxiliary matrix
                          \begin{equation}
                           {\bf S}_j^{(t+1)}=[{\bf S}_j^{(t)}, {\bf s}_j^{(m_j)}={\bf x}_{t+1}],
                        \end{equation}
                          whereas ${\bf S}_i^{(t+1)}={\bf S}_i^{(t)}$, for all $i\neq j$.
                     \item If $t>T_{train}$: update the parameters of $j$-th Gaussian, 
                       \begin{equation}
                         \label{EqUp1}
                               {\bm \mu}_j^{(t+1)}=\frac{1}{m_j}\sum_{i=1}^{m_j}{\bf s}_j^{(i)},
                       \end{equation}
                        and
                        \begin{gather}
                               \small
                        \begin{split}                   
                               \label{EqUp2}
                               {\bf C}_j^{(t+1)}=\frac{{\bf \tilde{S}}_j^{(t+1)} \cdot  \big[{\bf \tilde{S}}_j^{(t+1)}\big]^T +(m_j-1) \epsilon {\bf I}_d}{m_j-1}, 
                       \end{split} 
                       \end{gather}           
                           whrere  ${\bf \tilde{S}}_j^{(t+1)} = {\bf S}_j^{(t+1)} - {\bm \mu}_j^{(t+1)} \otimes {\bf 1}_{m_j}^{\top}$,
        with $\otimes$ denoting the Kronecker product \citep{VanLoan2000}. Morev,er set ${\bm \mu}_i^{(t+1)}={\bm \mu}_i^{(t)}$, ${\bf C}_i^{(t+1)}={\bf C}_i^{(t)}$, $\forall i\neq j$. Since $m_j$ has been incremented, then update also the weights              
                               \begin{equation}
                                         \label{EqUp3}
                                        w_i^{(t+1)}=\frac{m_i}{\sum_{k=1}^{N} m_k}, \quad i=1,...,N,
                               \end{equation}
                            so that ${\bf w}^{(t+1)}=[w_1^{(t+1)},...,w_N^{(t+1)}]^T$.                  
                            \end{enumerate}
 \item If $t<T_{tot}$ repeat from step 2.
\end{enumerate}
Observe that the proposal PDF is only updated for $T_{train}<t<T_{stop}$.
Moreover, note that the matrix ${\bf S}_i^{(t)}$ (and ${\bf \tilde{S}}_i^{(t)}$), for some $i\in \{1,...,N\}$, has dimension $d\times m_i$, so that ${\bf C}_i^{(t)}$ has always dimension $d\times d$.
The identity matrix  $\epsilon {\bf I}_d$ is used just to avoid numerical problems (the matrix ${\bf C}_i^{(t)}$ must be positive definite), as in \citep{Haario01}.


\section{Efficient Recursive Update of the Parameters}
\label{sec:recursive}
\label{UpdateParameter}
To update the parameters of the selected ($j$-th) Gaussian PDF in the mixture, we can use recursive expressions. Indeed, 
Recalling that $m_j=m_j+1$ is already updated and ${\bf s}_j^{(m_j)}={\bf x}_{t+1}$ in step \ref{Step3b} of the algorithm, \eqref{EqUp1} can be rewritten as 
            \begin{equation}
                         \label{EqUp1rec}
                               {\bm \mu}_j^{(t+1)}=\frac{1}{m_j} {\bf x}_{t+1}+\frac{m_j-1}{m_j}  {\bm \mu}_j^{(t)},
             \end{equation}   
          and the Eq. \eqref{EqUp2} becomes         
         \begin{gather}
         \small
            \begin{split}    
                         \label{EqUp2rec}
                              {\bf C}_j^{(t+1)}=\frac{1}{m_j-1} \Bigg[ &\frac{({\bf x}_{t+1}-{\bm \mu}_j^{(t+1)})({\bf x}_{t+1}-{\bm \mu}_j^{(t+1)})^T}{m_j}+\epsilon {\bf I}_d \Bigg] +\frac{m_j-2}{m_j-1}  {\bf C}_j^{(t)}.\\
             \end{split}
             \end{gather}
            Finally, note that 
            $$\sum_{k=1}^{N} m_k=t+1+N,$$ 
 for $T_{train}<t<T_{stop}$, so that 
                                   \begin{equation}
                                         \label{EqUp3rec}
                                                       w_i^{(t+1)}=\frac{m_i}{t+N+1}, \quad i=1,...,N,
                               \end{equation}
In this way, the novel technique becomes computationally efficient in high dimensional problems, as well.

\section{AGM-MH as black-box method}
\label{sec:blackBox}

The AGM-MH method shows sensitive dependence on the initial conditions.
If some prior information about the target is available, it can be used to choose the initial parameters.
If no prior informations is available, the AGM-MH can be applied as black-box algorithm in the following way: 
\begin{itemize}
	\item Use a great number $N$ of Gaussians (typically the number must be greater if the dimension $d$ increases).
	\item Select randomly the means ${\bm \mu_{1:N}^{(0)}}$ in order to cover as possible  the support domain $\mathcal{D} \subseteq 
		\mathbb{R}^d$. 
	\item Choose diagonal covariance matrices ${\bf C}_{1:N}^{(0)}=\sigma^2 {\bf I}_d$, with a big value of $\sigma^2$ in order to 
		explore the space $\mathcal{D} \subseteq \mathbb{R}^d$ in the train period, $t\leq T_{train}$.
	\item The parameter $T_{train}$ should be chosen bigger for greater dimensions $d$. Numerical results suggest that  $T_{train}=100d$ 
		could be a suitable choice. In general, for more complicated target distributions a greater $T_{train}$ could be needed.     
\end{itemize}
The use of a huge number of Gaussians does not generate computational problems since in the adaptation of the AGM-MH the weights of the irrelevant vanish quickly to zero. Therefore, the computational cost is controlled by the adaptation so that only the Gaussians located close to high probability regions survive. The useless Gaussian PDFs located faraway to the modes of the target are virtually discarded (the weights becomes quickly zero).   
Finally it is important to remark that, in general, the proposal is refined from the initial setting and the performance is improved.


\section{Simulations}
\label{sec:results}

\subsection{Example 1}
In this toy example, we apply the AGM-MH method to draw from a univariate bimodal target PDF defined as 
\begin{gather}
\begin{split}
p_o(x)\propto p(x)&=\exp\left\{-\frac{(x^2-4)^2}{4}\right\}\\
&=\exp\left\{-\frac{x^4-4x^2+16}{4}\right\},
\end{split}
\end{gather}
that, clearly, has two modes at $x=\pm 2$. We set $N=2$, number of Gaussian PDFs in the proposal PDF, $w_{i}^{(0)}=0.5$, and $(\sigma_i^2)^{(0)}=10$ with $i=1,2$. The two initial means $\mu_1^{(0)}\sim \mathcal{U}([-4,0])$, $\mu_1^{(0)}\sim \mathcal{U}([0,4])$  are chosen uniformly in $[-4,0]$ and $[0,4]$, respectively. We draw $T_{tot}=5000$ iterations of the chain, and set $T_{train}=200$, $T_{stop}=T_{tot}$ (i.e., the adaptation is never stopped).  The initial state is randomly choose as $x_{0}\sim N(x;0,1)$.    
We use {\it all} the generated samples to estimate the mean of the target (that is $0$ since $p(x)$ is symmetric).  
The mean square error (MSE) of the estimation (averaged over $2000$ runs) is $\approx 15 \cdot 10^{-4}$.  The estimated linear correlation between contiguous samples is $\approx 0.18$. Without using any adaptation, namely using a standard MH, the correlation is $\approx 0.78$. Hence, we can observe as the correlation decreases using the AGM-MH algorithm.

The final averaged locations of the means of the proposal are $\mu_1^{(T_{tot})}\approx -1.88$, $\mu_2^{(T_{tot})}\approx 1.88$. The final weights of the mixture are $w_{i}^{(T_{tot})}\approx 0.5$ and $\sigma_i^2\approx 0.16$, $i=1,2$.
\begin{figure*}[htb]
\centering 
\centerline{
 \subfigure[]{\includegraphics[width=6.5cm]{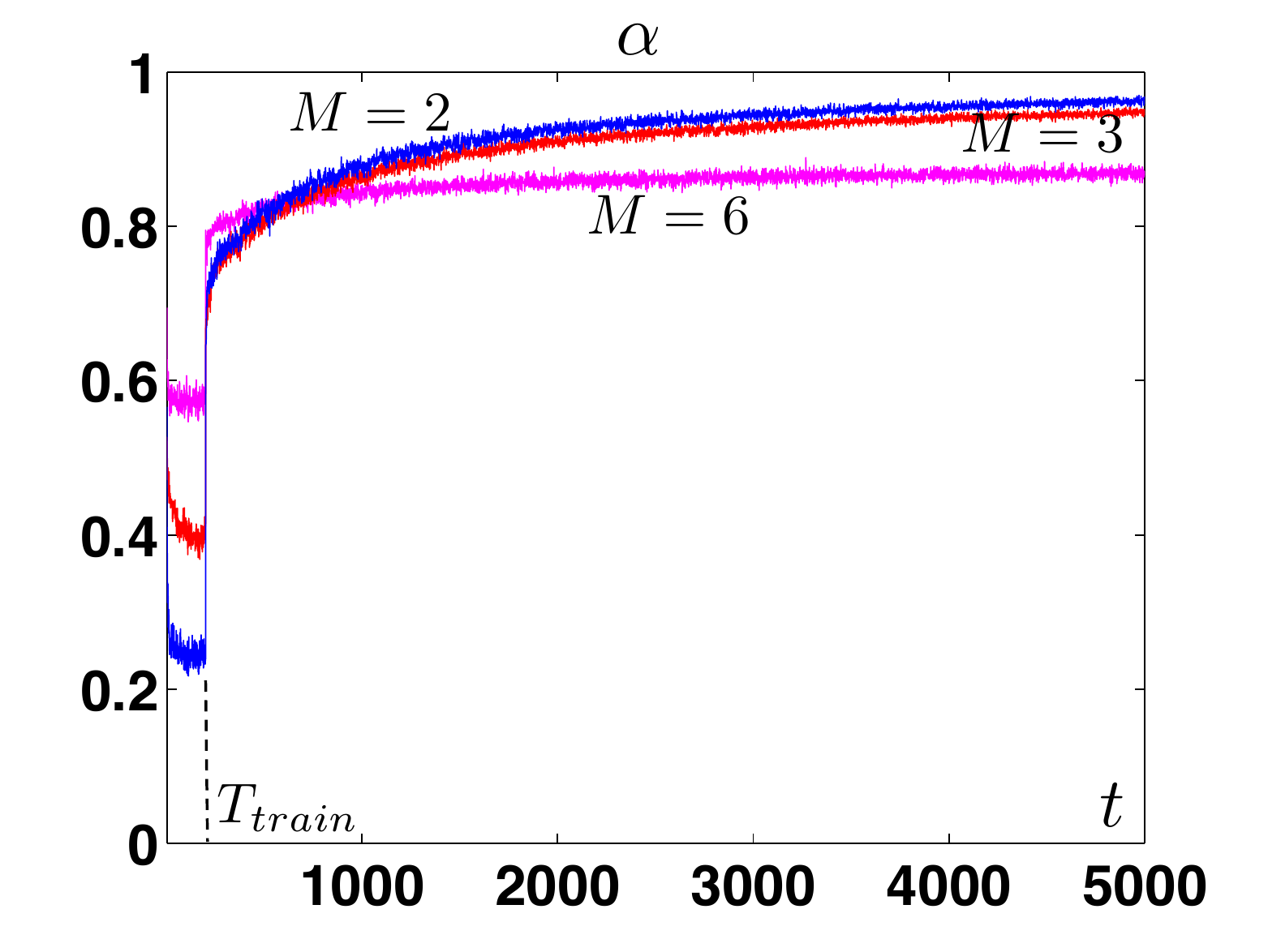}}
 }
\centerline{
 \subfigure[]{\includegraphics[width=6.5cm]{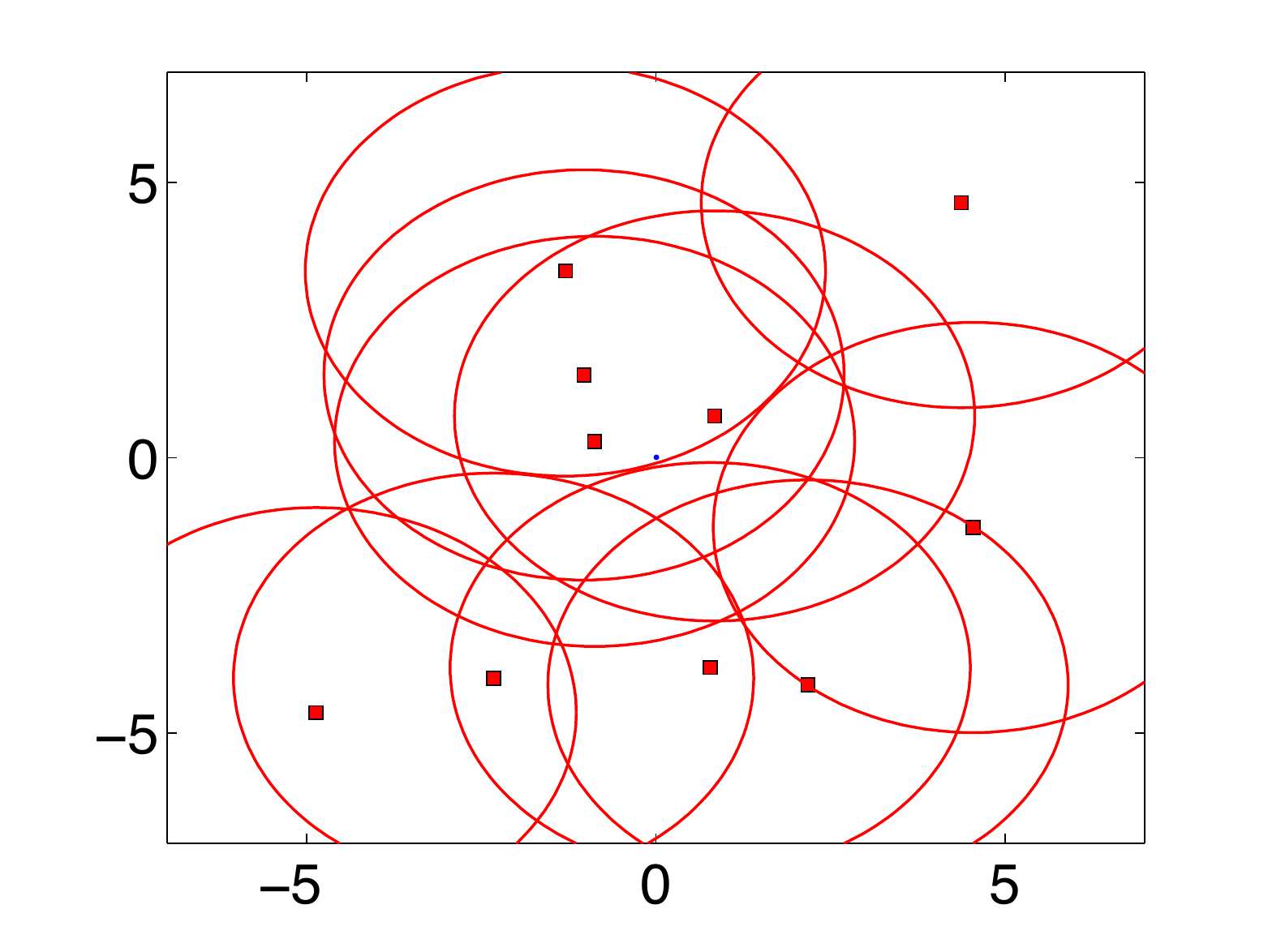}}
 \subfigure[]{\includegraphics[width=6.6cm]{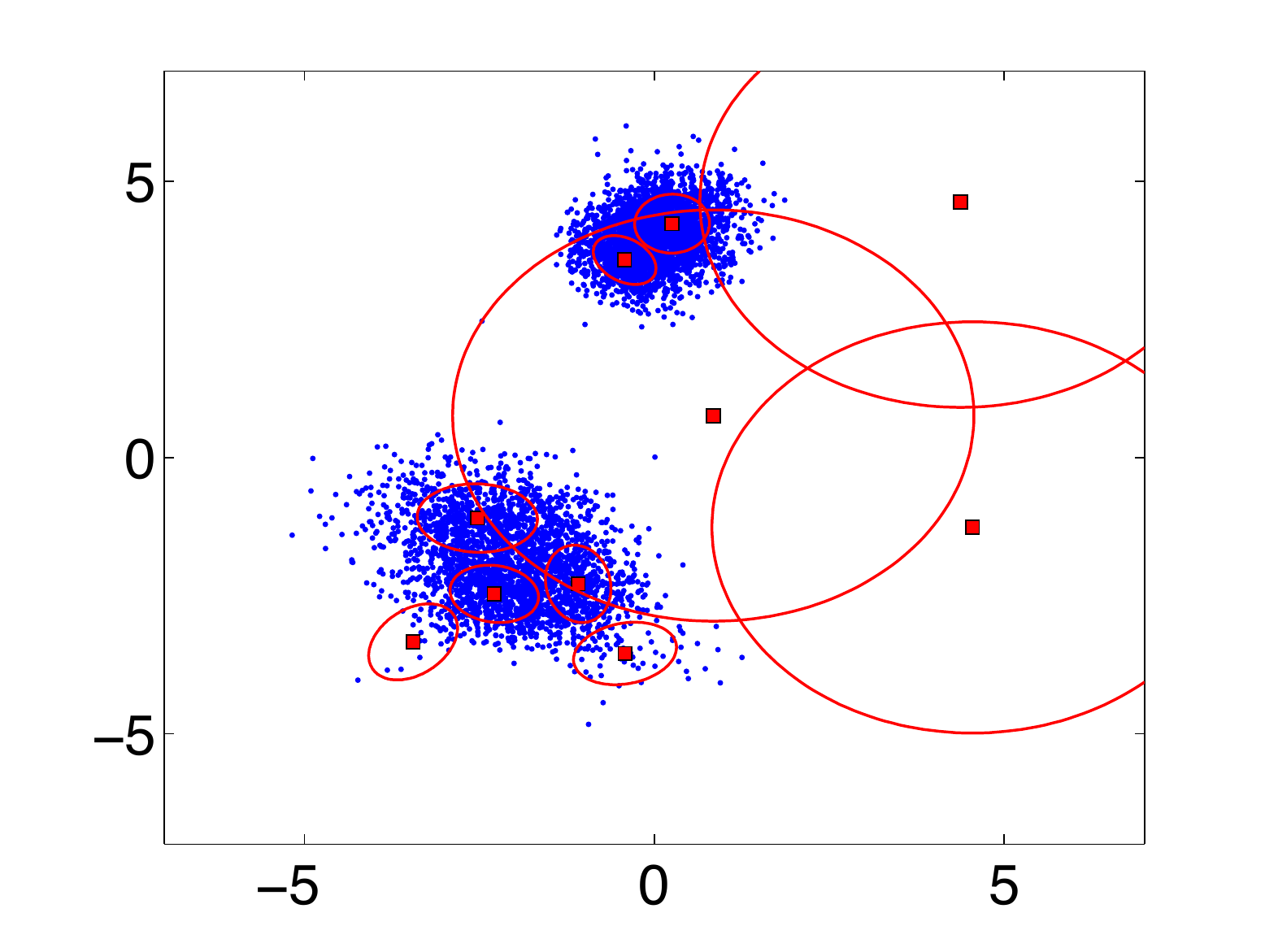}}
 }
  \caption{{\bf (a)} The averaged values of  $\alpha$ as function of the iteration index $t$ for different  $M=2,3,6$. For $t>T_{train}$, the $\alpha$ grows since the proposal is closer to the target. {\bf (b)} The initial configuration of the means (squares) and the covariance matrices (ellipses). {\bf (c)} The final configuration at $t=T_{tot}=7000$.  }
\label{fig1}
\end{figure*}

\subsection{Example 2}
For the sake of simplicity, we consider again an univariate target density. However, now we consider that target distribution is itself a mixture of Gaussian PDFs.\footnote{It is important to remark that we consider a mixture of Gaussians as a target PDF just to discuss the performance of the AGM-MH algorithm. More specifically, we desire to show as, in this case, the proposal convergences to real shape of the target, depending on the initial setting. However, clearly, the algorithm can be used to draw from any kind of target distribution.}
Specifically, the target PDF is formed by $M$-Gaussians, i.e.,
\begin{gather}
\begin{split}
p_o(x)\propto p(x)=\sum_{i=1}^{M}a_i \mathcal{N}(x|  \eta_i, \rho_i^2),
\end{split}
\end{gather}
where the weights are $a_i=1/M$ and the variances $\rho_i^2=4$, $i=1,...,M$.
We consider different $3$ cases with $M=2,3,6$. The means are, for each case, 
$$\eta_i=-10,10, \mbox{ }\mbox{ }\mbox{ for }\mbox{ }\mbox{ } M=2, $$
$$\eta_i=-10,0,10, \mbox{ }\mbox{ }\mbox{ for }\mbox{ }\mbox{ } M=3, $$
$$\eta_i=-15,-10,-5,5,10,15 \mbox{ }\mbox{ }\mbox{ for }\mbox{ }\mbox{ } M=6, $$
with $i=1,...,M$. 
In the proposal we also use $N=M$ Gaussians and each initial mean is chosen uniformly in $[-20,20]$. All the initial variances are set $(\sigma_j^2)^{(0)}=10$ and the weights $w_j^{(0)}=1/N$, $j=1,...,N$.   

As in the first example, we draw $T_{tot}=5000$ iterations of the chain, set $T_{train}=200$ and $T_{stop}=T_{tot}$ (i.e., the adaptation is never stopped). The initial state of the chain is randomly choose as $x_{0}\sim N(x;0,1)$. We use all the generated samples to estimate the normalizing constant of the target.  
The mean square error (MSE) of the estimation (averaged over $1000$ runs) is $1.6 \cdot 10^{-4}$, $1.1 \cdot 10^{-4}$ and $2 \cdot 10^{-5}$ for $M=2,3,6$ respectively.
The resulting correlation is $0.13$, $0.14$ and $0.16$ for $M=2,3,6$, respectively. With a standard MH (without adaptation) the correlation is $0.81$, $0.72$ and $0.46$,  for $M=2,3,6$. Hence, another time, we remark as the adaptation of the AGM-MH reduces considerably the correlation among the generated samples. 
Finally, Fig. \ref{fig1}(a) depicts the averaged values  of the acceptance function $\alpha$ in Eq. \eqref{AlphaEq} as function of $t$ and for different $M$. In this case, for $t>T_{train}$, the averaged values of $\alpha$ increase because of the proposal becomes closer to the target PDF owing to the adaptation.



\subsection{Example 3}
In this example, our goal is drawing from a bivariate target PDF using the AGM-MH as a black-box technique. Just for simplicity, we also consider a bivariate mixture of $M=2$ Gaussians as target distribution, with means ${\bm \eta}_1=[-2,-2]^{T}$, ${\bm \eta}_2=[0,4]^{T}$ and covariance matrices ${\bm \Sigma}_1=[0.3 \ \ 0.1; \ 0.1 \ \ 0.3]$, ${\bm \Sigma}_2=[0.8 \ \ -0.3; \ -0.3 \ \ 0.8]$. The weights are $a_i=0.5,0.5$, $i=1,2$.     
We set  $T_{tot}=7000$, set $T_{train}=200$ and $T_{stop}=T_{tot}$ (i.e., the adaptation is never stopped).
First, we use for the proposal $N=2$ Gaussian PDFs, $w_i^{(0)}=0.5$, ${\bf C}_i^{(0)}=10{\bf I}_d$ for $i=1,2$. The means are selected uniformly, ${\bm \mu}_1\sim \mathcal{U}([-5,5]\times[0,5])$ and ${\bm \mu}_2\sim \mathcal{U}([-5,5]\times[-5,0])$. In this case, all the parameters of the mixture in the proposal convergences always to the corresponding values of the mixture forming the target PDF. 

Moreover, we also consider the case with $N=10$. We choose the means ${\bm \mu}_i$ of the proposal uniformly in the square $[-5,5]\times [-5,5]$. In this situation, the AGM-MH refines the initial proposal PDF improving the parameters of the Gaussians in the mixture that are in good locations, whereas the weights of the unhelpful Gaussians decrease quickly to zero and their parameters remain invariant, as shown in Fig. \ref{fig1}.
\section{Discussion}
\label{sec:discussion}

We have proposed a novel adaptive independent MH algorithm (AGM-MH) to draw samples from arbitrary multi-modal and multi-dimensional targets.
AGM-MH builds on the work of \citep{Haario01}, extending it by using a Gaussian mixture proposal and updating also the means and the weights of the Gaussians.
Compared to a previous extension provided by \citep{Giordani10}, our approach is more efficient, updates the proposal at every iteration instead of only at a fixed number of iterations.

\bibliography{bibliografia,refsAGM}

\begin{thebibliography}{13}
\providecommand{\natexlab}[1]{#1}
\providecommand{\url}[1]{\texttt{#1}}
\expandafter\ifx\csname urlstyle\endcsname\relax
  \providecommand{\doi}[1]{doi: #1}\else
  \providecommand{\doi}{doi: \begingroup \urlstyle{rm}\Url}\fi

\bibitem[Liu(2004)]{Liu04b}
J.~S. Liu.
\newblock \emph{{M}onte {C}arlo Strategies in Scientific Computing}.
\newblock Springer, 2004.

\bibitem[Liang et~al.(2010)Liang, Liu, and Caroll]{Liang10}
F.~Liang, C.~Liu, and R.~Caroll.
\newblock \emph{Advanced {M}arkov {C}hain {M}onte {C}arlo Methods: Learning
  from Past Samples}.
\newblock Wiley Series in Computational Statistics, England, 2010.

\bibitem[Robert and Casella(2004)]{Robert04}
C.~P. Robert and G.~Casella.
\newblock \emph{{M}onte {C}arlo Statistical Methods}.
\newblock Springer, 2004.

\bibitem[Wang et~al.(2002)Wang, Chen, and Liu]{Wang2002}
Xiaodong Wang, Rong Chen, and Jun~S. Liu.
\newblock Monte {C}arlo {B}ayesian signal processing for wireless
  communications.
\newblock \emph{Journal of VLSI Signal Processing}, 30:\penalty0 89--105, 2002.

\bibitem[Andrieu et~al.(2003)Andrieu, {De Freitas}, Doucet, and
  Jordan]{Andrieu2003}
Christophe Andrieu, Nando {De Freitas}, Arnaud Doucet, and Michael~I. Jordan.
\newblock An introduction to {MCMC} for machine learning.
\newblock \emph{Machine Learning}, 50:\penalty0 5--43, 2003.

\bibitem[Fitzgerald(2001)]{Fitzgerald01}
W.~J. Fitzgerald.
\newblock {M}arkov chain {M}onte {C}arlo methods with applications to signal
  processing.
\newblock \emph{Signal Processing}, 81\penalty0 (1):\penalty0 3--18, January
  2001.

\bibitem[Metropolis et~al.(1953)Metropolis, Rosenbluth, Rosenbluth, Teller, and
  Teller]{Metropolis53}
N.~Metropolis, A.~Rosenbluth, M.~Rosenbluth, A.~Teller, and E.~Teller.
\newblock Equations of state calculations by fast computing machines.
\newblock \emph{Journal of Chemical Physics}, 21:\penalty0 1087--1091, 1953.

\bibitem[Hastings(1970)]{Hastings70}
W.~K. Hastings.
\newblock {M}onte {C}arlo sampling methods using {M}arkov chains and their
  applications.
\newblock \emph{Biometrika}, 57\penalty0 (1):\penalty0 97--109, 1970.

\bibitem[Martino and M\'{\i}guez(2010)]{MartinoSigPro10}
L.~Martino and J.~M\'{\i}guez.
\newblock Generalized rejection sampling schemes and applications in signal
  processing.
\newblock \emph{Signal Processing}, 90\penalty0 (11):\penalty0 2981--2995,
  November 2010.

\bibitem[Martino et~al.(2012)Martino, Read, and Luengo]{MartinoA2RMS}
L.~Martino, J.~Read, and D.~Luengo.
\newblock Improved adaptive rejection {M}etropolis sampling algorithms.
\newblock \emph{arXiv:1205.5494}, May 2012.

\bibitem[Haario et~al.(2001)Haario, Saksman, and Tamminen]{Haario01}
H.~Haario, E.~Saksman, and J.~Tamminen.
\newblock An adaptive {M}etropolis algorithm.
\newblock \emph{Bernoulli}, 7\penalty0 (2):\penalty0 223--242, April 2001.

\bibitem[Giordani and Kohn(2010)]{Giordani10}
P.~Giordani and R.~Kohn.
\newblock Adaptive independent {M}etropolis-{H}astings by fast estimation of
  mixtures of normals.
\newblock \emph{Journal of Computational and Graphical Statistics}, 19\penalty0
  (2):\penalty0 243--259, September 2010.

\bibitem[{Van Loan}(2000)]{VanLoan2000}
Charles~F. {Van Loan}.
\newblock The ubiquitous {K}ronecker product.
\newblock \emph{Journal of Computational and Applied Mathematics},
  123:\penalty0 85--100, 2000.

\end{thebibliography}

\end{document}